\documentclass[twocolumn,prb,amsmath,amssymb,floatfix]{revtex4}

\usepackage{bbold}
\usepackage{bm}
\usepackage{color}
\usepackage{dcolumn}
\usepackage{feynmf} 
\usepackage{graphics}
\usepackage{graphicx}
\usepackage{subfigure}
\usepackage{slashed}
\usepackage{placeins}

\def\al{\alpha}

\def\veps{\varepsilon}

\def\be{\begin{equation}}
\def\ee{\end{equation}}
\def\bea{\begin{eqnarray}}
\def\eea{\end{eqnarray}}
\def\bse{\begin{subequations}}
\def\ese{\end{subequations}}
\def\bc{\begin{center}}
\def\ec{\end{center}}

\def\ms{\medskip}

\def\ni{\noindent}
\def\ra{\rightarrow}

\def\nonum{\nonumber}

\def\D{{\rm d}}
\def\Ord{{\rm O}}

\newcommand{\comment}[1]{}

\catcode`,\active

\catcode`\,12

\begin{document}

\title{Note on an application of the method of uniqueness \\ to reduced quantum electrodynamics}

\author{A.~V.~Kotikov$^1$ and S. Teber$^2$}
\affiliation{
$^1$Bogoliubov Laboratory of Theoretical Physics, Joint Institute for Nuclear Research, 141980 Dubna, Russia.\\
$^2$Laboratoire de Physique Th\'eorique et Hautes Energies, Universit\'e Pierre et Marie Curie, 4 place Jussieu, 75005, Paris, France.}

\date{\today}

\begin{abstract}
Using the method of uniqueness a two-loop massless propagator Feynman diagram 
with a noninteger index on the central line is evaluated in a very 
transparent way. The result is applied to the computation of the two-loop
polarization operator in reduced quantum electrodynamics.
\end{abstract}

\maketitle

\begin{fmffile}{fmfIlambda}

{\sl Introduction} - The exact analytical computation of multi-loop Feynman diagrams is of crucial importance for the evaluation
of renormalization group functions, {\it i.e.}, $\beta$ functions and anomalous dimensions of fields.
Since the early days of quantum field theory a variety of methods have been developed, and often combined, in order to achieve this task, 
{\it e.g.}, the Gegenbauer polynomial technique~\cite{CKT-80,Kotikov-96}, integration by parts~\cite{VPK-81,CT-81}, and
the method of uniqueness~\cite{VPK-81,Usyukina-83,Kazakov-84,Kazakov-85}. 
The latter allows, in principle, the computation of complicated Feynman diagrams using sequences of simple transformations.
A given diagram is straightforwardly integrated once the appropriate sequence is found. The task of finding such a sequence for a given diagram is, however,
nontrivial, see Ref.~[\onlinecite{Kazakov-lecture}] for a review.

\begin{figure}[ht]
   \includegraphics{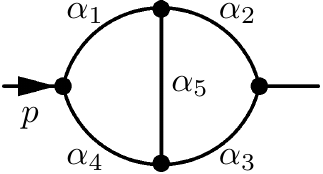}
   \caption{\label{fig:J}
   Two-loop massless propagator diagram.}
\end{figure}

One of the basic building blocks of multiloop calculations is the two-loop massless propagator diagram:
\bea
&&J(\al_1,\al_2,\al_3,\al_4,\al_5) = 
\label{def:J}
\\
&&\int \int \frac{\D^D k_1 \, \D^D k_2}{k_1^{2\al_1}\,k_2^{2\al_2}\,(k_2-p)^{2\al_3}\,(k_1-p)^{2\al_4}\,(k_2-k_1)^{2\al_5}} \, ,
\nonum
\eea
with arbitrary indices $\al_i$ and external momentum $p$ in a Euclidean space-time of dimensionality $D$, Fig.~\ref{fig:J}, 
see Ref.~[\onlinecite{Grozin}] for a historical review on this diagram.
When all indices are integers the diagram of Eq.~(\ref{def:J}) is well known and easily calculated.
Its evaluation for arbitrary indices is however highly nontrivial:
the results can be represented~\cite{Bierenbaum} as a combination 
of twofold series. In some particular cases, however, the results can be 
obtained~\cite{VPK-81,Kazakov-85,Gracey-92,KSV-93,Kotikov-96,BGK-97,BK-98} in 
significantly simpler form.
In Ref.~[\onlinecite{Kotikov-96}] a class of complicated diagrams, with two integer indices on adjacent lines 
and three other arbitrary indices, has been computed exactly on the basis of a new development of the GP technique. 
The latter approach is technically involved and the result is expressed in terms of a generalized hypergeometric function, ${}_3F_2$ with argument $1$. 
For this class of diagrams, similar results have been found in Ref.~[\onlinecite{BGK-97}] using an ansatz to solve the
recurrence relations for the two-loop diagram arising from IBP. 
In this brief report we consider the simplest but important diagram belonging to this class, whose coefficient function reads
\be
I(\lambda) = \frac{p^{2(2-\lambda)}}{\pi^D}\,J(1,1,1,1,\lambda), \quad \lambda=\frac{D}{2}-1 \, ,
\label{def:I}
\ee

\FloatBarrier

\begin{figure}[ht]
    \includegraphics{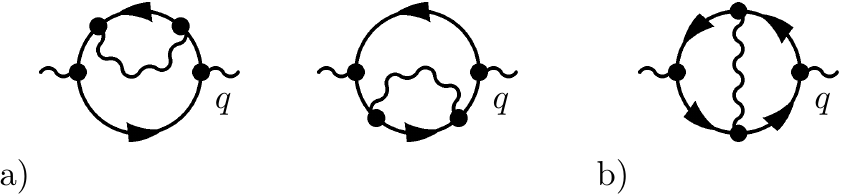}
    \caption{\label{fig:rqed-2loop-polarization}
     Two-loop vacuum polarization diagrams.}
\end{figure}
\FloatBarrier

\ni where the index $\al_5$ has been restricted to $\lambda$, all other indices being $1$.
This diagram has been already calculated\cite{VPK-81} (see also discussions in Ref.~[\onlinecite{KSV-93}]) but it seems that our evaluation
is simpler and more transparent.
As will be seen in the following, the diagram can indeed be straightforwardly integrated
using the method of uniqueness in momentum space with the help of a simple but ingenious three-step transformation.
As an application, we compute the two-loop polarization operator (Fig.~\ref{fig:rqed-2loop-polarization})
in reduced quantum electrodynamics~\cite{Gorbar2001}.

\ms

{\sl Definitions and notations} - In what follows we use dimensional regularization. All calculations are performed in a Euclidean space-time of dimensionality
$D=2+2\lambda$ which may either be even dimensional ($\lambda \ra 1$) or odd dimensional ($\lambda \ra 1/2$). 
Diagrams will be analyzed in momentum space. For a given diagram, the integrations are over 
loop momenta and the lines are simple power laws of the form: $1/k^{2\al}$ where $\al$ is the index of the line. 
The index of a diagram is defined as the sum of the indices of its constituent lines. 
A line with an arbitrary index $\al$ can be represented graphically as:
\be
\parbox{10mm}{
  \begin{fmfgraph*}(10,10)
    \fmfleft{i}
    \fmfright{o}
    \fmf{fermion,label=$\al$,l.s=left}{i,o}
  \end{fmfgraph*}
} \, \Rightarrow \frac{1}{k^{2\al}}\,.
\label{def:index}
\ee
In order to simplify notations, arrows will not be displayed in the following but they are implied, {\it e.g.}, Fig.~\ref{fig:J}.

\begin{widetext}
With these definitions and graphical notations, chains reduce to the product of propagators:
\be
\parbox{15mm}{
  \begin{fmfgraph*}(15,10)
    \fmfleft{i}
    \fmfright{o}
    \fmf{plain,label=$\al$,l.s=left}{i,v}
    \fmf{plain,label=$\beta$,l.s=left}{v,o}
    \fmfdot{v}
  \end{fmfgraph*}
} \, = \,
\parbox{15mm}{
  \begin{fmfgraph*}(15,10)
    \fmfleft{i}
    \fmfright{o}
    \fmf{plain,label=$\al+\beta$,l.s=left}{i,o}
  \end{fmfgraph*}
}\,\quad . 
\label{def:chain}
\ee
On the other hand simple loops involve an integration:
\bea
\parbox{15mm}{
  \begin{fmfgraph*}(15,13)
      \fmfleft{i}
      \fmfright{o}
      \fmf{plain,right=0.6,label=$\al$,l.d=0.1h}{i,o}
      \fmf{plain,right=0.6,label=$\beta$,l.d=0.05h}{o,i}
      \fmfdot{i,o}
  \end{fmfgraph*}
} \,\,\, = \, \pi^{D/2} A(\al,\beta)\quad
\parbox{15mm}{
  \begin{fmfgraph*}(15,10)
    \fmfleft{i}
    \fmfright{o}
    \fmf{plain,label=$\al+\beta-D/2$,l.s=left}{i,o}
    \fmfdot{i,o}
  \end{fmfgraph*}
}\quad, \qquad
A(\al,\beta) = \frac{a(\al)a(\beta)}{a(\al+\beta-D/2)}, \quad a(\al) = \frac{\Gamma(D/2-\al)}{\Gamma(\al)} \, .
\label{def:loop}
\eea
In momentum space a vertex and a triangle are said to be unique if their indices are equal to $D/2=1+\lambda$ and
$D=2+2\lambda$, respectively. They are related to each other by the uniqueness (or star-triangle) 
relation:
\be
\parbox{16mm}{
  \begin{fmfgraph*}(16,16)
    \fmfleft{l}
    \fmfleft{vl}
    \fmfright{r}
    \fmfright{vr}
    \fmftop{t}
    \fmftop{vt}
    \fmffreeze
    \fmfforce{(-0.1w,0.042265h)}{l}
    \fmfforce{(0.1w,0.2h)}{vl}
    \fmfforce{(1.1w,0.042265h)}{r}
    \fmfforce{(0.9w,0.2h)}{vr}
    \fmfforce{(.5w,1.1h)}{t}
    \fmfforce{(.5w,0.966025h)}{vt}
    \fmffreeze
    \fmf{plain,label=$\al_2$,label.side=left,l.d=0.05w}{vr,vl}
    \fmf{plain,label=$\al_3$,label.side=left,l.d=0.05w}{vl,vt}
    \fmf{plain,label=$\al_1$,label.side=left,l.d=0.05w}{vt,vr}
    \fmf{plain}{l,vl}
    \fmf{plain}{r,vr}
    \fmf{plain}{t,vt}
    \fmffreeze
    \fmfdot{vl,vt,vr}
  \end{fmfgraph*}
} \qquad \underset{\underset{i}{\sum} \al_i = D}{=} \qquad \pi^{D/2} A(\al_1,\al_2)\,
\parbox{16mm}{
  \begin{fmfgraph*}(16,16)
    \fmfleft{l}
    \fmfright{r}
    \fmftop{t}
    \fmfleft{v}
    \fmffreeze
    \fmfforce{(-0.1w,0.142265h)}{l}
    \fmfforce{(1.1w,0.142265h)}{r}
    \fmfforce{(.5w,1.1h)}{t}
    \fmfforce{(.5w,0.488675h)}{v}
    \fmffreeze
    \fmf{plain,label=$\tilde{\al}_2$,label.side=left,l.d=0.05w}{v,t}
    \fmf{plain,label=$\tilde{\al}_3$,label.side=right,l.d=0.05w}{r,v}
    \fmf{plain,label=$\tilde{\al}_1$,label.side=left,l.d=0.05w}{v,l}
    \fmffreeze
    \fmfdot{v}
  \end{fmfgraph*}
}\, \qquad ,
\label{def:star-triangle}
\ee
where $\tilde{\al}_i=D/2-\al_i$ is the index dual to $\al_i$. 
Finally, for an arbitrary triangle (unique or not) the following recurrence relation is obtained from integration by parts~\cite{VPK-81,CT-81}: 
\bea
(D-\al_2-\al_3-2\al_5) \quad
\parbox{16mm}{
    \begin{fmfgraph*}(16,14)
      \fmfleft{i}
      \fmfright{o}
      \fmfleft{ve}
      \fmfright{vo}
      \fmftop{vn}
      \fmftop{vs}
      \fmffreeze
      \fmfforce{(-0.1w,0.5h)}{i}
      \fmfforce{(1.1w,0.5h)}{o}
      \fmfforce{(0w,0.5h)}{ve}
      \fmfforce{(1.0w,0.5h)}{vo}
      \fmfforce{(.5w,0.95h)}{vn}
      \fmfforce{(.5w,0.05h)}{vs}
      \fmffreeze
      \fmf{plain}{i,ve}
      \fmf{plain,left=0.8}{ve,vo}
      \fmf{phantom,left=0.5,label=$\al_1$,l.d=-0.01w}{ve,vn}
      \fmf{phantom,right=0.5,label=$\al_2$,l.d=-0.01w}{vo,vn}
      \fmf{plain,left=0.8}{vo,ve}
      \fmf{phantom,left=0.5,label=$\al_3$,l.d=-0.01w}{vo,vs}
      \fmf{phantom,right=0.5,label=$\al_4$,l.d=-0.01w}{ve,vs}
      \fmf{plain,label=$\al_5$,l.d=0.05w}{vs,vn}
      \fmf{plain}{vo,o}
      \fmffreeze
      \fmfdot{ve,vn,vo,vs}
    \end{fmfgraph*}
}
\quad & = & \,\, \al_2 \left[\quad
\parbox{16mm}{
    \begin{fmfgraph*}(16,14)
      \fmfleft{i}
      \fmfright{o}
      \fmfleft{ve}
      \fmfright{vo} 
      \fmftop{vn}
      \fmftop{vs}
      \fmffreeze
      \fmfforce{(-0.1w,0.5h)}{i}
      \fmfforce{(1.1w,0.5h)}{o}
      \fmfforce{(0w,0.5h)}{ve}
      \fmfforce{(1.0w,0.5h)}{vo}
      \fmfforce{(.5w,0.95h)}{vn}
      \fmfforce{(.5w,0.05h)}{vs}
      \fmffreeze
      \fmf{plain}{i,ve}
      \fmf{plain,left=0.8}{ve,vo}
      \fmf{phantom,left=0.5}{ve,vn}
      \fmf{phantom,right=0.4,label=$+$,l.d=-0.01w}{vo,vn}
      \fmf{plain,left=0.8}{vo,ve}
      \fmf{phantom,left=0.5}{vo,vs}
      \fmf{phantom,right=0.5}{ve,vs}
      \fmf{plain,label=$-$,l.d=0.05w}{vs,vn}
      \fmf{plain}{vo,o}
      \fmffreeze
      \fmfdot{ve,vn,vo,vs}
    \end{fmfgraph*}
} \qquad - \qquad
\parbox{16mm}{
    \begin{fmfgraph*}(16,14)
      \fmfleft{i}
      \fmfright{o}
      \fmfleft{ve}
      \fmfright{vo} 
      \fmftop{vn}
      \fmftop{vs}
      \fmffreeze
      \fmfforce{(-0.1w,0.5h)}{i}
      \fmfforce{(1.1w,0.5h)}{o}
      \fmfforce{(0w,0.5h)}{ve}
      \fmfforce{(1.0w,0.5h)}{vo}
      \fmfforce{(.5w,0.95h)}{vn}
      \fmfforce{(.5w,0.05h)}{vs}
      \fmffreeze
      \fmf{plain}{i,ve}
      \fmf{plain,left=0.8}{ve,vo}
      \fmf{phantom,left=0.4,label=$-$,l.d=-0.01w}{ve,vn}
      \fmf{phantom,right=0.4,label=$+$,l.d=-0.01w}{vo,vn}
      \fmf{plain,left=0.8}{vo,ve}
      \fmf{phantom,left=0.5}{vo,vs}
      \fmf{phantom,right=0.5}{ve,vs}
      \fmf{plain}{vs,vn}
      \fmf{plain}{vo,o}
      \fmffreeze
      \fmfdot{ve,vn,vo,vs}
    \end{fmfgraph*}
} \quad
\right]
\nonum
\\ &\qquad& \nonum 
\\ & + & \,\,
\al_3 \left[\quad
\parbox{16mm}{
    \begin{fmfgraph*}(16,14)
      \fmfleft{i}
      \fmfright{o}
      \fmfleft{ve}
      \fmfright{vo}
      \fmftop{vn}
      \fmftop{vs}
      \fmffreeze
      \fmfforce{(-0.1w,0.5h)}{i}
      \fmfforce{(1.1w,0.5h)}{o}
      \fmfforce{(0w,0.5h)}{ve}
      \fmfforce{(1.0w,0.5h)}{vo}
      \fmfforce{(.5w,0.95h)}{vn}
      \fmfforce{(.5w,0.05h)}{vs}
      \fmffreeze
      \fmf{plain}{i,ve}
      \fmf{plain,left=0.8}{ve,vo}
      \fmf{phantom,left=0.5}{ve,vn}
      \fmf{phantom,right=0.5}{vo,vn}
      \fmf{plain,left=0.8}{vo,ve}
      \fmf{phantom,left=0.4,label=$+$,l.d=-0.01w}{vo,vs}
      \fmf{phantom,right=0.5}{ve,vs}
      \fmf{plain,label=$-$,l.d=0.05w}{vs,vn}
      \fmf{plain}{vo,o}
      \fmffreeze
      \fmfdot{ve,vn,vo,vs}
    \end{fmfgraph*}
} \qquad - \qquad
\parbox{16mm}{
    \begin{fmfgraph*}(16,14)
      \fmfleft{i}
      \fmfright{o}
      \fmfleft{ve}
      \fmfright{vo}
      \fmftop{vn}
      \fmftop{vs}
      \fmffreeze
      \fmfforce{(-0.1w,0.5h)}{i}
      \fmfforce{(1.1w,0.5h)}{o}
      \fmfforce{(0w,0.5h)}{ve}
      \fmfforce{(1.0w,0.5h)}{vo}
      \fmfforce{(.5w,0.95h)}{vn}
      \fmfforce{(.5w,0.05h)}{vs}
      \fmffreeze
      \fmf{plain}{i,ve}
      \fmf{plain,left=0.8}{ve,vo}
      \fmf{phantom,left=0.5}{ve,vn}
      \fmf{phantom,right=0.5}{vo,vn}
      \fmf{plain,left=0.8}{vo,ve}
      \fmf{phantom,left=0.4,label=$+$,l.d=-0.01w}{vo,vs}
      \fmf{phantom,right=0.4,label=$-$,l.d=-0.01w}{ve,vs}
      \fmf{plain}{vs,vn}
      \fmf{plain}{vs,vn}
      \fmf{plain}{vo,o}
      \fmffreeze
      \fmfdot{ve,vn,vo,vs}
    \end{fmfgraph*}
} \quad
\right]\, ,
\label{def:IBP}
\eea
where $\pm$ on the right-hand side of the equation denotes the increase or decrease of a line index by $1$ with respect to its value on the left-hand side.
In the following, in order to simply notations, we will assume that lines with no index are ordinary lines. In momentum space ordinary lines have index
$\al=1$.

\ms

{\sl Calculation of the diagram} - With the help of the above notations and identities we proceed on calculating $I(\lambda)$. 
The first transformation consists in replacing the central line by a 
loop\footnote{In coordinate space, it corresponds to the insertion of a 
point into this line (see the table of such transformations in 
Ref.~[\onlinecite{VPK-81}] and also Ref.~[\onlinecite{Kazakov-lecture}] for
a review).}, Eq.~(\ref{def:loop}), in order
to make the right triangle unique. The uniqueness relation, Eq.~(\ref{def:star-triangle}), can then be used. 
In graphical notations this reads
\be
J(1,1,1,1,\lambda) \,\, = \quad
\parbox{18mm}{
  \begin{fmfgraph*}(18,16)
    \fmfleft{i}
    \fmfright{o}
    \fmfleft{ve}
    \fmfright{vo}
    \fmftop{vn}
    \fmftop{vs}    
    \fmffreeze
    \fmfforce{(-0.1w,0.5h)}{i}
    \fmfforce{(1.1w,0.5h)}{o}
    \fmfforce{(0w,0.5h)}{ve}
    \fmfforce{(1.0w,0.5h)}{vo}
    \fmfforce{(.5w,0.95h)}{vn}
    \fmfforce{(.5w,0.05h)}{vs}
    \fmffreeze
    \fmf{plain}{i,ve}
    \fmf{plain,left=0.8}{ve,vo}
    \fmf{plain,left=0.8}{vo,ve}
    \fmf{plain,label=$\lambda$,l.d=0.05w}{vs,vn}
    \fmf{plain}{vo,o}
    \fmffreeze
    \fmfdot{ve,vn,vo,vs}
  \end{fmfgraph*}
} \quad  =  \, \frac{1}{\pi^{D/2} A(1,2\lambda)} \quad
\parbox{18mm}{
  \begin{fmfgraph*}(18,16)
    \fmfleft{i}
    \fmfright{o}
    \fmfleft{ve}
    \fmfright{vo}
    \fmftop{vn}
    \fmftop{vs}
    \fmffreeze
    \fmfforce{(-0.1w,0.5h)}{i}
    \fmfforce{(1.1w,0.5h)}{o}
    \fmfforce{(0w,0.5h)}{ve}
    \fmfforce{(1.0w,0.5h)}{vo}
    \fmfforce{(.5w,0.95h)}{vn}
    \fmfforce{(.5w,0.05h)}{vs}
    \fmffreeze
    \fmf{plain}{i,ve}
    \fmf{plain,left=0.8}{ve,vo}
    \fmf{plain,left=0.8}{vo,ve}
    \fmf{plain,left=0.3}{vs,vn}
    \fmf{plain,left=0.3,label=$2\lambda$,l.d=0.05w}{vn,vs}
    \fmf{plain}{vo,o}
    \fmffreeze
    \fmfdot{ve,vn,vo,vs}
  \end{fmfgraph*}
} \quad = \quad
\parbox{16mm}{
  \begin{fmfgraph*}(16,16)
    \fmfleft{i}
    \fmfright{o}
    \fmfleft{ve}
    \fmfright{vo}
    \fmftop{vn}
    \fmftop{vs}
    \fmffreeze
    \fmfforce{(-0.1w,0.5h)}{i}
    \fmfforce{(1.1w,0.5h)}{o}
    \fmfforce{(0w,0.5h)}{ve}
    \fmfforce{(1.0w,0.5h)}{vo}
    \fmfforce{(.5w,0.9h)}{vn}
    \fmfforce{(.5w,0.1h)}{vs}
    \fmffreeze
    \fmf{plain}{i,ve}
    \fmf{plain,left=0.8}{ve,vo}
    \fmf{plain,left=0.8}{vo,ve}
    \fmf{plain}{vs,vn}
    \fmf{phantom,right,label=$\lambda$,l.s=left,l.d=-0.05h}{vo,vn}
    \fmf{phantom,left,label=$\lambda$,l.s=right,l.d=-0.05h}{vo,vs}
    \fmf{plain}{vo,o}
    \fmffreeze
    \fmfdot{ve,vn,vo,vs}
  \end{fmfgraph*}
} \quad \frac{1}{p^{2(1-\lambda)}} \, .
\label{first+second-steps}
\ee
Finally, using integration by parts, Eq.~(\ref{def:IBP}), the last diagram is reduced to sequences of chains and 
simple loops which can immediately be integrated with the help of Eqs.~(\ref{def:chain}) and
(\ref{def:loop}):
\begin{subequations}
\label{third-step}
\bea
( -2 \delta ) \quad
\parbox{16mm}{
  \begin{fmfgraph*}(16,16)
    \fmfleft{i}
    \fmfright{o}
    \fmfleft{ve}
    \fmfright{vo}
    \fmftop{vn}
    \fmftop{vs}
    \fmffreeze
    \fmfforce{(-0.1w,0.5h)}{i}
    \fmfforce{(1.1w,0.5h)}{o}
    \fmfforce{(0w,0.5h)}{ve}
    \fmfforce{(1.0w,0.5h)}{vo}
    \fmfforce{(.5w,0.9h)}{vn}
    \fmfforce{(.5w,0.1h)}{vs}
    \fmffreeze
    \fmf{plain}{i,ve}
    \fmf{plain,left=0.8}{ve,vo}
    \fmf{plain,left=0.8}{vo,ve}
    \fmf{plain}{vs,vn}
    \fmf{phantom,right=0.1,label=$\lambda+\delta$,l.s=right}{vo,vn}
    \fmf{phantom,left=0.1,label=$\lambda+\delta$,l.s=left}{vo,vs}
    \fmf{plain}{vo,o}
    \fmffreeze
    \fmfdot{ve,vn,vo,vs}
  \end{fmfgraph*}
} \, \, &=& \,\, 2 ( \lambda + \delta ) \quad 
\left[ \quad
\parbox{32mm}{
  \begin{fmfgraph*}(32,16)
    \fmfleft{i}
    \fmfright{o}
    \fmfleft{ve}
    \fmfright{vo}
    \fmftop{v}
    \fmffreeze
    \fmfforce{(-0.1w,0.5h)}{i}
    \fmfforce{(1.1w,0.5h)}{o}
    \fmfforce{(0w,0.5h)}{ve}
    \fmfforce{(1.0w,0.5h)}{vo}
    \fmfforce{(.5w,0.5h)}{v}
    \fmffreeze
    \fmf{plain}{i,ve}
    \fmf{plain,left=0.8}{ve,v}
    \fmf{plain,left=0.8}{v,ve}
    \fmf{plain,left=0.8,label=$\lambda+\delta$,l.s=left}{v,vo}
    \fmf{plain,left=0.8,label=$\lambda+\delta+1$,l.s=left}{vo,v}
    \fmf{plain}{vo,o}
    \fmffreeze
    \fmfdot{ve,v,vo}
  \end{fmfgraph*}
} \qquad - \qquad
\parbox{16mm}{
  \begin{fmfgraph*}(16,16)
    \fmfleft{i}
    \fmfright{o}
    \fmfleft{ve}
    \fmfright{vo}
    \fmftop{v}
    \fmffreeze
    \fmfforce{(-0.1w,0.5h)}{i}
    \fmfforce{(1.1w,0.5h)}{o}
    \fmfforce{(0w,0.5h)}{ve}
    \fmfforce{(1.0w,0.5h)}{vo}
    \fmfforce{(.5w,0.9h)}{v}
    \fmffreeze
    \fmf{plain}{i,ve}
    \fmf{plain,left=0.8}{ve,vo}
    \fmf{plain,left=0.8,label=$\lambda+\delta+1$,l.s=left}{vo,ve}
    \fmf{plain,left=0.5}{v,ve}
    \fmf{phantom,right=0.1,label=$\lambda+\delta$,l.s=right}{vo,v}
    \fmf{plain}{vo,o}
    \fmffreeze
    \fmfdot{ve,v,vo}
  \end{fmfgraph*}
} \qquad
\right] \,\, ,
\label{third-step-a}\\
&\quad&
\nonumber \\
&\quad&
\nonumber \\
& = & \frac{\pi^D 2 ( \lambda + \delta )}{p^{2(1+2\delta)}}\, A(1,1)\, \Big[ A(\lambda+\delta+1,\lambda+\delta) - A(\lambda+\delta+1, 1+\delta) \Big]_1 \, ,
\label{third-step-b}
\eea
\end{subequations}
where the parameter $\delta$ has been introduced. The function $A(1,1)$ and the bracketed terms in Eq.~(\ref{third-step-b}) read
\begin{subequations}
\label{A(1,1) and first-bracket}
\bea
A(1,1) & = & \frac{\Gamma^2(\lambda)\Gamma(1-\lambda)}{\Gamma(2\lambda)} \, ,
\label{A(1,1)} \\ 
( \lambda + \delta ) \, \Big[ \bullet \Big]_1   & = &  \frac{\Gamma(-\delta)}{\Gamma(\lambda+\delta)}\, \frac{\Gamma(\lambda-\delta)\Gamma(1+2\delta)}{\Gamma(1+\delta)\Gamma(\lambda-2\delta)}\,
\Bigg[\, \frac{\Gamma(1-\delta) \Gamma(1+\delta) \Gamma(\lambda+2\delta) \Gamma(\lambda-2\delta)}{\Gamma(1-2\delta) \Gamma(1+2\delta)\Gamma(\lambda+\delta) \Gamma(\lambda-\delta)} - 1 \,\Bigg]_2 \, .
\label{first-bracket}
\eea
\end{subequations}
At this point it is convenient to use the following product expansion of the Gamma function:
\be
\Gamma(x+\veps) = \Gamma(x)\,\exp \Big[ \,\,\sum_{k=1}^{\infty} \psi^{(k-1)}(x) \frac{\veps^k}{k!} \,\,\Big]\, ,
\quad
\psi(x) = \psi^{(0)}(x) = \frac{\Gamma'(x)}{\Gamma(x)}, \quad  \psi^{(k)}(x) = \frac{\D^k}{\D x^k} \,\psi(x)\, ,
\label{Gamma-expansion}
\ee
where $\psi^{(k)}$ is the polygamma function of order $k$.
From Eq.~(\ref{Gamma-expansion}), the following relation is obtained:
\be
\Gamma(x+\veps) \Gamma(x-\veps) = \Gamma^2(x)\,\exp \Big[\,\,2\, \sum_{m=1}^{\infty} \psi^{(2m-1)}(x) \frac{\veps^{2m}}{(2m)!}\,\, \Big]\,.
\label{Gamma-expansion-2}
\ee
Making use of Eq.~(\ref{Gamma-expansion-2}) in the bracket of Eq.~(\ref{first-bracket}) yields
\be
\Big[ \bullet \Big]_2 =  \exp \Big[ \,\, 2\,\sum_{m=1}^{\infty} \left( 2^{2m}-1 \right) \left[\psi^{(2m-1)}(\lambda) - \psi^{(2m-1)}(1)\right] \frac{\delta^{2m}}{(2m)!} \,\,\Big]
= 3 \delta^2 \Big[ \psi'(\lambda) - \psi'(1) \Big]_3 + \Ord(\delta^4)\,.
\label{second-bracket}
\ee
Substituting back Eq.~(\ref{second-bracket}) in (\ref{A(1,1) and first-bracket}) and performing the remaining $\delta$ expansion yields
\be
( \lambda + \delta ) \, \Big[ \bullet \Big]_1 = \frac{-3 \delta}{\Gamma(\lambda)}\,\Big[ \bullet \Big]_3 \,\, 
\underset{{\rm Eq.}~(\ref{third-step})}{\Rightarrow} \quad 
\parbox{16mm}{
  \begin{fmfgraph*}(16,16)
    \fmfleft{i}
    \fmfright{o}
    \fmfleft{ve}
    \fmfright{vo}
    \fmftop{vn}
    \fmftop{vs}
    \fmffreeze
    \fmfforce{(-0.1w,0.5h)}{i}
    \fmfforce{(1.1w,0.5h)}{o}
    \fmfforce{(0w,0.5h)}{ve}
    \fmfforce{(1.0w,0.5h)}{vo}
    \fmfforce{(.5w,0.9h)}{vn}
    \fmfforce{(.5w,0.1h)}{vs}
    \fmffreeze
    \fmf{plain}{i,ve}
    \fmf{plain,left=0.8}{ve,vo}
    \fmf{plain,left=0.8}{vo,ve}
    \fmf{plain}{vs,vn}
    \fmf{phantom,right=0.1,label=$\lambda$,l.s=right}{vo,vn}
    \fmf{phantom,left=0.1,label=$\lambda$,l.s=left}{vo,vs}
    \fmf{plain}{vo,o}
    \fmffreeze
    \fmfdot{ve,vn,vo,vs}
  \end{fmfgraph*}
} \quad = \, \frac{\pi^D}{p^2}\,3\,\frac{\Gamma(\lambda)\Gamma(1-\lambda)}{\Gamma(2\lambda)} \,\Big[ \psi'(\lambda) - \psi'(1) \Big] \,,
\label{third-bracket}
\ee
where, in the last step, Eq.~(\ref{third-step}) has been used and $\delta$ sent to zero. 
Substituting the final result of Eq.~(\ref{third-bracket})
in Eq.~(\ref{first+second-steps}) and using Eq.~(\ref{def:I}), we obtain the advertised result\cite{VPK-81,KSV-93} for the coefficient function:
\be
I(\lambda) = 3\,\frac{\Gamma(\lambda)\Gamma(1-\lambda)}{\Gamma(2\lambda)} \,\Big[ \psi'(\lambda) - \psi'(1) \Big]\,,
\label{result:Ilambda}
\ee
where $\psi'(x)$ is the trigamma function.
In the even-dimensional case ($\lambda \ra 1$ or $D \ra 4$) the well-known result $I(1) = 6 \, \zeta(3)$, is obtained.
On the other hand, in the odd-dimensional case ($\lambda \ra 1/2$ or $D \ra 3$), which is one of the cases of interest to Ref.~[\onlinecite{ST2012,KSV-93}], 
the result reads $I(1/2) = 6 \pi \,\zeta(2)$.

\ms

{\sl Application} - We now focus on the computation of radiative corrections to the polarization operator 
$\Pi^{\mu\nu}(q) = \Pi(q^2) \, (g^{\mu\nu} q^2 - q^{\mu}q^{\nu})$
in reduced quantum electrodynamics~\cite{Gorbar2001} (RQED$_{d_\gamma,d_e}$). The latter
describes the interaction of a photon field living in $d_\gamma$ dimensions with a fermion field living in a reduced space-time of $d_e$ dimensions ($d_e \leq d_\gamma$).
Within dimensional regularization, the computation of Feynman integrals in such a reduced theory can be carried out by introducing two epsilon parameters,
$\veps_\gamma$ and $\veps_e$, such that $d_\gamma = 4-2\veps_\gamma$ and $d_e=4-2\veps_e-2\veps_\gamma$, respectively.
In Ref.~[\onlinecite{ST2012}] the corrections up to two loops (see Fig.~\ref{fig:rqed-2loop-polarization} where the corresponding diagrams were displayed) 
were computed for an arbitrary RQED$_{d_\gamma,d_e}$ using the general result of Ref.~[\onlinecite{Kotikov-96}] for $I(\lambda)$. The resulting expression is rather cumbersome.
Here, we focus on the case of RQED$_{4,d_e}$. In the limit $\veps_\gamma \ra 0$ and using Eq.~(\ref{result:Ilambda}) we obtain the following simpler and more explicit formulas
(see definitions in Ref.~[\onlinecite{ST2012}]):
\begin{subequations}
\label{2loopdiagrams-ab}
\bea
2\Pi_{2a}(q^2) = && d\,
\frac{e^4 \, \Gamma(\lambda) \Gamma^2(1+\varepsilon_\gamma)}{
(4\pi)^{3+\lambda- 2\varepsilon_\gamma}
\left(q^2\right)^{1-\lambda+2\varepsilon_\gamma}}\, 
\frac{16 \Gamma(1+\lambda)
 \Gamma(1-\lambda)}{ \Gamma(3+2\lambda)}\, \nonumber \\
&& \times \biggl\{ \, \lambda^2 \left(\frac{1}{\varepsilon_\gamma} + 
\overline{\psi} + \frac{2}{1+2\lambda}\right)
+ \frac{3\lambda}{2} - 2 + \frac{2}{1+\lambda} \, + \, \Ord(\veps_\gamma) \, \Biggr\} \, ,
\label{2loop-a} \\
\Pi_{2b}(q^2) = && d\,
\frac{e^4 \, \Gamma(\lambda) \Gamma^2(1+\varepsilon_\gamma)}{
(4\pi)^{3+\lambda- 2\varepsilon_\gamma}
\left(q^2\right)^{1-\lambda+2\varepsilon_\gamma}}\, 
\frac{16 \Gamma(1+\lambda)
 \Gamma(1-\lambda)}{ \Gamma(3+2\lambda)} \nonumber \\
&& \times \biggl\{\, -\lambda^2 \left(\frac{1}{\varepsilon_\gamma} + 
\overline{\psi} + \frac{2}{1+2\lambda}\right)
+ \frac{\lambda}{2} - \frac{1}{2} - \frac{3}{2\lambda} - \frac{1}{1+\lambda}
 + \frac{3}{2} \lambda (1+\lambda)
\Bigr[ \psi'(\lambda) - \psi'(1) \Bigl] \, +  \,\Ord(\veps_\gamma) \, \Biggr\} \,,
\label{2loop-b}
\eea
\end{subequations}
where $\lambda=1-\veps_e$
and $\overline{\psi} = 3\psi(2\lambda) - 2 \psi(\lambda) + 2\psi(1-\lambda) -3 \psi(1)$.
The one-loop and total two-loop contributions therefore read
\begin{subequations}
\label{1+2loopdiagrams}
\bea
\Pi_{1}(q^2) =&& - d\,\frac{e^2 \, \Gamma(\lambda)}{(4\pi)^{1+\lambda} \, \left(q^2\right)^{1-\lambda}}
\frac{\Gamma(1+\lambda) \Gamma(1-\lambda)}{ (1+2\lambda) \Gamma(2\lambda)} \, , 
\label{1loop}\\
\Pi_{2}(q^2) = && d\,
\frac{e^4 \, \Gamma(\lambda)}{(4\pi)^{3+\lambda} \, \left(q^2\right)^{1-\lambda}}
\frac{16 \Gamma(1+\lambda) \Gamma(1-\lambda)}{ \Gamma(3+2\lambda)} \, C_1(\lambda) \, , 
\label{2loop-a+b} \\
C_1(\lambda) = && 2\lambda - \frac{5}{2} - \frac{3}{2\lambda} + \frac{1}{1+\lambda}
+ \frac{3}{2} \lambda (1+\lambda) \Bigl[ \psi'(\lambda) - \psi'(1) \Bigr] \, .
\label{C1}
\eea
\end{subequations}
From Eqs.~(\ref{1+2loopdiagrams}) we see that $\Pi_{1}(q^2)$ and $\Pi_{2}(q^2)$ are finite as long as $\lambda \neq 1$.
We can then replace $e^2$ by $4\pi \alpha$ in  (\ref{1+2loopdiagrams}) which yields
\be
\label{2loopdiagramN}
\Pi_{2}(q^2) =  - \frac{\alpha}{\pi\lambda(1+\lambda)} \, C_1(\lambda) \, 
\Pi_{1}(q^2) \, .
\ee
It should be noted that these results cannot be used for QED$_{4}$
which can be reached from a general RQED$_{d_\gamma,d_e}$
(see Ref.~[\onlinecite{ST2012}]) by first fixing $\veps_e=0$ and then
taking the limit $\varepsilon_\gamma \to 0$. The results (\ref{1+2loopdiagrams}) 
are singular in the limit $\lambda \to 1$ but this limit corresponds to $\varepsilon_\gamma = 0$ and $\varepsilon_e \to 0$, which does
not lead to QED$_4$.

The total, up to two loops, gauge-field self-energy  in RQED$_{4,d_e}$
($\varepsilon_\gamma = 0$ and arbitrary $\varepsilon_e$)
may then be written as
\bea
\label{1+2loopdiagram}
\Pi(q^2) &&= \Pi_{1}(q^2) \Bigl(1+\alpha C(\lambda) + \Ord \left( {\bf \alpha}^2\right) 
\Bigr), \nonumber \\
~~ C(\lambda) &&= - \frac{1}{\pi\lambda(1+\lambda)} \, C_1(\lambda)
= - \frac{1}{2\pi} \left(3 \Bigl[ \psi'(\lambda+2) - \psi'(1) \Bigr]
+ \frac{4}{1+\lambda} + \frac{1}{(1+\lambda)^2} \right)
 \, .
\eea
\end{widetext}
For $\lambda=1/2$, {\it i.e.}, in the case of RQED$_{4,3}$ ($\varepsilon_\gamma = 0$ and 
$\varepsilon_e=1/2$) which corresponds to an ultrarelativistic model of graphene~\cite{GrapheneRG} (a 2-brane), 
we reproduce the basic result of Ref.~[\onlinecite{ST2012}] 
:~\footnote{A coefficient similar to $C_1(1/2)$ has been obtained earlier in [\onlinecite{GusyninHR00}] 
in the framework of the $1/N$ expansion of QED$_{3}$. The similarity of the coefficients
comes from the corresponding similarity of the effective photon propagators
in RQED$_{4,3}$ (see Eq. (9) in [\onlinecite{ST2012}]) and in the $1/N$ expansion of
QED$_{3}$ (see Eq. (2.16) in [\onlinecite{AppelquistBKW86}]) at large $N$.}
\be
\label{C+C1}
C_1(1/2)= \frac{9\pi^2-92}{24},~~ C(1/2)= \frac{92-9\pi^2}{18\pi} \,.
\ee
This coefficient is small, $C(1/2) \approx 0.056$, in qualitative agreement with some results obtained in the nonrelativistic limit~\cite{C-non-relat-small} 
(see however Refs.~[\onlinecite{C-non-relat-large}]) as well as experimental results~\cite{C-experiment} where $C(1/2)$ corresponds to an interaction correction 
coefficient to the optical conductivity of undoped graphene.

As a next step of our future investigations we would like to evaluate the fermion self-energy
of general RQED$_{d_\gamma,d_e}$ in the ultrarelativistic limit.

\acknowledgments
We are grateful to Andrey Grozin and Valery Gusynin for useful discussions.
The work of A.V.K.\ was supported in part by the Russian Foundation for Basic
Research (Grant No. 13-02-01005).


\end{fmffile}
\end{document}